\documentclass[journal]{IEEEtran}
\usepackage{amsmath,amsfonts}
\usepackage{algorithmic}
\usepackage{algorithm}
\usepackage{array}
\usepackage[caption=false,font=normalsize,labelfont=sf,textfont=sf]{subfig}
\usepackage{textcomp}
\usepackage{stfloats}
\usepackage{url}
\usepackage[hidelinks]{hyperref}
\usepackage{verbatim}
\usepackage{graphicx}
\usepackage{cite}
\usepackage{booktabs} 
\usepackage{multirow} 
\usepackage{xcolor}
\newcommand{\greencheck}{{\color{green}\checkmark}}
\newcommand{\redtimes}{{\color{red}$\times$ }}

\begin{document}

\title{IoT- and AI-informed urban air quality models for vehicle pollution monitoring}
\author{Jan M. Armengol$^{a,b}$,
Vicente  Masip$^{b}$,
Ada Barrantes$^{b}$,
Gabriel M. Beltrami$^{b}$,
Sergi Albiach$^{b}$,
Daniel Rodriguez-Rey$^{b,c}$,
Marc Guevara$^{b}$,
Albert Soret$^{b}$,
Eduardo Quiñones$^{b}$,
Elli Kartsakli$^{b}$
\thanks{$^{a}$Department of Fluid Mechanics, Universitat Politècnica de Catalunya $\cdot$ BarcelonaTech (UPC), Barcelona, Spain}
\thanks{$^{b}$Barcelona Supercomputing Center, Barcelona, Spain}
\thanks{$^{c}$Now at Tecnalia, Basque Research and Technology Alliance (BRTA), Donostia, Spain}
}



\maketitle

\begin{abstract}
With the rise of intelligent Internet of Things (IoT) systems in urban environments, new opportunities are emerging to enhance real-time environmental monitoring. While most studies focus either on IoT-based air quality sensing or physics-based modeling in isolation, this work bridges that gap by integrating low-cost sensors and AI-powered video-based traffic analysis with high-resolution urban air quality models. We present a real-world pilot deployment at a road intersection in Barcelona’s Eixample district, where the system captures dynamic traffic conditions and environmental variables, processes them at the edge, and feeds real-time data into a high-performance computing (HPC) simulation pipeline. Results are validated against official air quality measurements of nitrogen dioxide (NO$_2$). Compared to traditional models that rely on static emission inventories, the IoT-assisted approach enhances the temporal granularity of urban air quality predictions of traffic-related pollutants. Using the full capabilities of an IoT-edge-cloud-HPC architecture, this work demonstrates a scalable, adaptive, and privacy-conscious solution for urban pollution monitoring and establishes a foundation for next-generation IoT-driven environmental intelligence.
\end{abstract}


\section{Introduction}

Air pollution in urban environments remains a major public health concern, primarily due to sustained exposure to traffic-related pollutants such as nitrogen dioxide (NO$_2$) and particulate matter (PM) \cite{who2021who}. The design of effective mitigation strategies requires first an accurate characterization of pollution levels across space and time. Official air quality monitoring stations (AQMS) provide reliable real-time measurements of pollutant concentrations and serve as the primary reference for urban air quality assessment. However, their high installation and maintenance costs limit their number, ranging from none in small to medium-sized cities to only a few dozen in larger metropolitan areas. Furthermore, AQMS typically offer limited spatial coverage, restricting their ability to fully capture intraurban variability. To address these gaps, low-cost sensors (LCSs) have emerged as a promising complementary solution, enabling wider spatial deployment and denser monitoring networks. However, LCS generally suffer from lower accuracy and remain limited to point-based measurements. 

Complementing these observational approaches, \textbf{urban air quality (AQ) systems} (combining emission inventories, meteorological data, and atmospheric transport processes) can fill spatial and temporal gaps in observations and support scenario analysis for informed decision making and policy development \cite{rodriguez2022extent}. Thus, urban AQ systems have become essential tools for air pollution management. However, their precision is often limited by persistent uncertainties, primarily stemming from emission inventories and the complex dynamics of urban atmospheric flows.  

The rapid development of Internet of Things (IoT) technologies has significantly increased the availability of real-time observational data, offering new opportunities to enhance the responsiveness and granularity of urban AQ systems. In parallel, edge computing enables scalable, energy-efficient solutions by shifting computation closer to data sources, thus reducing latency and improving system responsiveness. These capabilities allow for real-time processing of large volumes of environmental and traffic data using big data analytics and Artificial Intelligence (AI) to extract valuable knowledge for urban AQ systems. In this context, we highlight observational data of two different types to improve current urban AQ systems:

\begin{itemize}
    \item \textbf{Traffic video data}, which can improve the estimation of emission rates.  Most current emission models are based on averaged traffic flows \cite{guevara2020hermesv3}. Traffic cameras combined with computer vision algorithms can deliver much richer information, including vehicle classification by type, fine-grained movement analytics (e.g., speed and acceleration), and the detection of congestion or atypical traffic events: all of which are critical for accurately modeling traffic-related emissions.
    
    \item  \textbf{Environmental sensor data}, which can be leveraged to constrain the outputs of the AQ model using data assimilation and data fusion techniques \cite{criado2023data}. IoT enables the deployment of a heterogeneous ecosystem of sensors including AQ measurements, but also meteorological variables such as wind speed and direction, temperature, and humidity. Such observational data are key for improving and validating LCS and model predictions.
    
\end{itemize}

The technologies discussed above lay the foundation for a next-generation, IoT-driven approach to urban environmental monitoring. However, most existing studies focus either on IoT-based AQ monitoring \cite{rodulfo2020smart,cetinkaya2021distributed,mahajan2023ubiquitous} or on urban air pollution modelling \cite{martin2024using} in isolation. This work bridges that gap by demonstrating an integrated system in which IoT environmental sensing informs a physics-based urban AQ model executed on high-performance computing (HPC) infrastructure across a unified IoT–edge–cloud–HPC continuum. The approach is validated through a real-world pilot deployment at a road intersection in Barcelona (Spain). Specifically, we:

\begin{itemize}
\item describe the deployed IoT platform and the dataset it generates, including pollution concentration time series and road user information inferred from traffic cameras;
\item define and evaluate the physics-based urban air pollution modeling framework in the HPC environment;
\item demonstrate the improvement in the spatial and temporal representation of pollutant concentrations achieved by integrating model outputs with observational data.
\end{itemize}

Section II provides an overview of the technological foundations and the challenges addressed. Section III presents real-world results from a pilot study in Barcelona, validating the proposed architecture. Finally, Sections IV and V discuss generalizability of open issues and lessons learned.

\section{Methods and technical challenges}

\begin{figure*}[ht]
    \centering
    \includegraphics[width=1\textwidth,trim= 0 3cm 0 0 cm,clip]{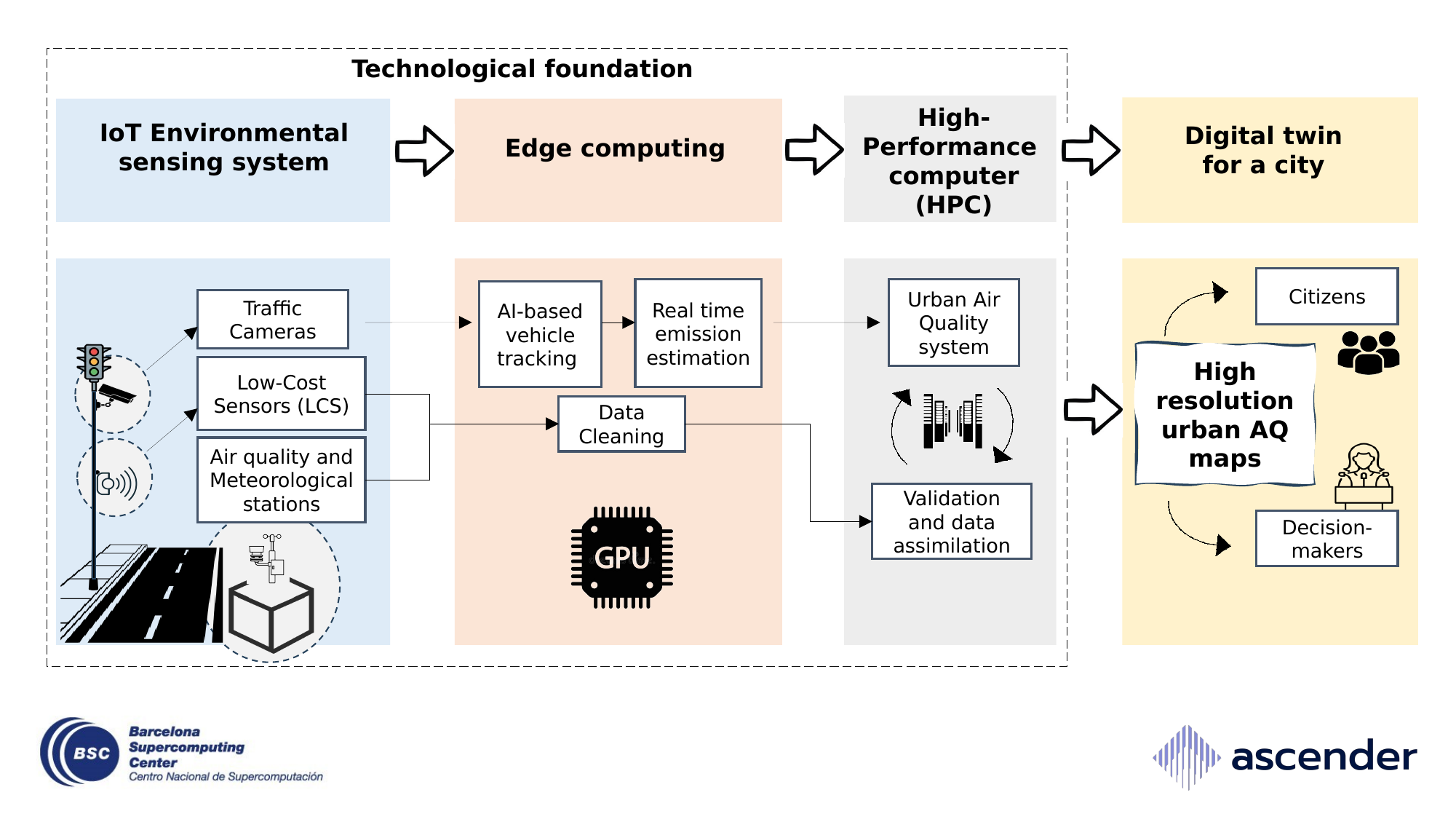}
    \caption{Data flow across the IoT-enabled air quality monitoring framework, showing how raw sensor and video data are processed at the edge, integrated into HPC-based urban simulations, and visualized through a cloud platform for decision-making and public communication.}
    \label{fig:workflow}
\end{figure*}

The system architecture outlined in Fig. \ref{fig:workflow} is organized into four distinct layers.
The first layer comprises the IoT data acquisition devices: (i) traffic cameras, (ii) low-cost sensor (LCS) networks, and (iii) high-precision reference stations. Data from these devices are transmitted to edge computing units via Ethernet and cellular connections (4G/5G).

At the edge computing layer, advanced tracking algorithms and vehicular emission models are executed to extract meaningful information from traffic videos. At the same time, calibration algorithms are applied to LCS data, and quality checks are performed on reference station measurements. The processed information is then transmitted to the HPC infrastructure for large-scale simulation and analysis, enabling parallel and scalable execution.

Within the HPC layer, real-time emission data provide dynamic inputs
for urban AQ simulations. At the same time, air pollution and meteorological observations serve two purposes: (i) data assimilation, in which observational data are merged with model outputs to correct biases and enhance accuracy, and (ii) model validation, to quantify output uncertainty.

This paper focuses on these three first layers, which together form the technological foundation of IoT-enabled urban AQ systems. 
Nevertheless, we also highlight in Fig. \ref{fig:workflow} a final cloud layer, where results are made accessible through an interactive dashboard to support decision-making and communicate real-time AQ information to citizens, with the potential to integrate scenario analysis through digital twins in future implementations.

Using IoT and advanced AQ modelling, this work presents innovations that address the following core challenges:
\begin{itemize}
    \item \textit{Challenge 1:} The high heterogeneity of edge, cloud and HPC infrastructures in terms of computing power, processor architectures and networking capabilities adds considerable complexity in the application development and deployment, thus hindering the creation of innovative complex data analytics services.

    \item \textit{Challenge 2:} The high volumes of generated data obtained by geographically disperse data sources require distributed edge computing and storage approaches to reduce latency and energy consumption.

    \item \textit{Challenge 3:} Privacy issues are a crucial concern when dealing with traffic surveillance cameras, as these systems can capture identifiable information such as vehicle license plates and pedestrian faces. The potential for re-identification poses ethical and legal challenges related to data protection, consent, and compliance with privacy regulations.

    \item \textit{Challenge 4:} The high computational demand of physics-based models represents a major challenge for developing operational urban air AQ systems, particularly when employing Computational Fluid Dynamics (CFD) to resolve pollutant dispersion at street or neighborhood scales. These simulations require substantial processing power and memory resources, which limit their ability to operate in real time. Achieving near-real-time performance becomes even more complex when accounting for rapidly changing meteorological conditions and dynamic emission patterns derived from IoT data. Therefore, efficient computational strategies are essential to ensure timely and scalable AQ model execution within an operational framework.
    
\end{itemize}

\section{Application to the Barcelona Pilot Case}

\begin{figure*}[ht]
    \centering
    \includegraphics[width=0.98\textwidth,trim= 3cm 2.5cm 2cm 2.5cm,clip]{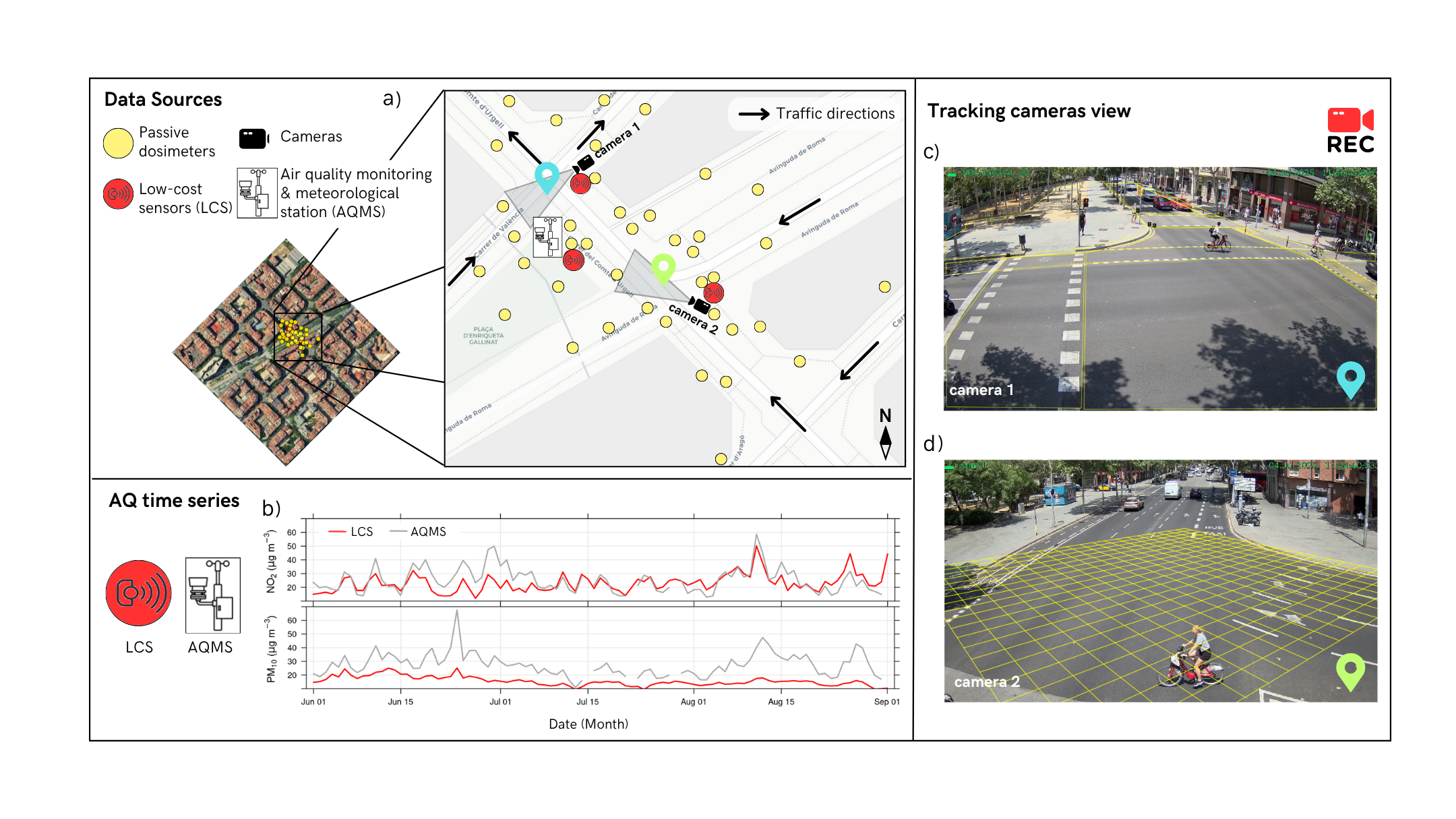}
    \caption{Overview of the pilot site and data sources. (a) Map of the intersection showing the deployed IoT sensing infrastructure, including traffic cameras, low-cost sensors, passive dosimeters, and the reference air quality monitoring station. (b) Comparison of NO$_2$ and PM$10$ time series measured by the low-cost sensors and the reference station. (c–d) Views from the two installed traffic cameras covering the monitored area.}
    \label{fig:pilot}
\end{figure*}

The city of Barcelona, located on the northeast coast of Spain, faces persistent NO$_2$ pollution due to high vehicle density and its compact urban morphology. The pilot study was conducted at a road intersection in the Eixample district, a centrally located area characterized by its grid-like layout and chamfered corners, which hosts dense commercial, cultural, and social activity. Fig. \ref{fig:pilot}a shows a map of the intersection and the deployed IoT sensing system. This site was selected for its combination of high traffic intensity, limited green space, and poor air quality, making it a representative and challenging environment for evaluating the proposed IoT-based air quality monitoring innovations. The following results illustrate the performance and insights gained from deploying the system in this real-world urban context.

\subsection{IoT environmental sensing system}

Two traffic cameras were installed on the traffic lights indicated in Fig. \ref{fig:pilot}a, making use of the physical structure and electrical access points available. The corresponding camera images, showing their monitored areas, are presented in Figs. \ref{fig:pilot}c and \ref{fig:pilot}d. The cameras were connected to GPU-enabled edge computing units located in a street cabinet via Ethernet connections.

\begin{table*}[ht]
\caption{Measured pollutants, meteorological variables, and temporal resolution of the environmental sensing system deployed in the Barcelona pilot.}
\label{tab:sensors}
\begin{tabular}{lccccccccccc}
\toprule
\multirow{2}{*}{\textbf{System}} & \multicolumn{7}{c}{\textbf{Air pollutants}} & \multicolumn{3}{c}{\textbf{Meteorology}} & \multirow{2}{*}{\textbf{Temporal Resolution}} \\
\cmidrule(lr){2-8} \cmidrule(lr){9-11}
 & NO & NO\textsubscript{2} & NO\textsubscript{x} & PM1 & PM2.5 & PM10 & CO/CO2 & Air Temperature & Relative Humidity & Wind &  \\
\midrule
LCS & \greencheck  & \greencheck & \redtimes & \greencheck & \greencheck & \greencheck & \greencheck & \greencheck & \greencheck & \redtimes & 1-minute averages \\
AQMS & \greencheck & \greencheck & \greencheck & \greencheck & \greencheck & \greencheck & \redtimes & \greencheck & \greencheck & \greencheck & 10-minute averages \\
Passive dosimeter & \redtimes & \greencheck & \redtimes & \redtimes & \redtimes & \redtimes & \redtimes & \redtimes & \redtimes & \redtimes & 4-week average \\
\hline
\\
\end{tabular}
\footnotesize{ Note: NO (nitric oxide), NO\textsubscript{2} (nitrogen dioxide), NO\textsubscript{x} (sum of reactive nitrogen species), PM1 (particles $<$1$\mu m$), PM2.5 ($<$2.5$\mu m$), PM10 ($<$10$\mu m$), CO (carbon monoxide), and CO\textsubscript{2} (carbon dioxide).}\\
\end{table*}

Reference air quality  data were obtained from an official AQ monitoring station (AQMS) of the Catalan monitoring network (XVPCA) \footnote{\href{https://mediambient.gencat.cat/ca/05_ambits_dactuacio/atmosfera/qualitat_de_laire/avaluacio/xarxa_de_vigilancia_i_previsio_de_la_contaminacio_atmosferica_xvpca/}{Xarxa de Vigilància i Previsió de la Contaminació Atmosfèrica (XVPCA).}} operated by the Catalan Regional Administration. This AQMS is equipped with an anemometer that measures wind speed and direction at 6.2 m above ground.  To complement these reference measurements, low-cost sensors (LCS) were deployed at each camera location, with one additional sensor co-located at the AQMS for calibration and validation. Prior to their final deployment, all three LCS were co-located with the AQMS for a period of five months. The co-location data were used to calibrate regression models that adjust for local environmental conditions. In addition to the IoT sensors, seven passive dosimeter campaigns were conducted during 2024 and 2025. These campaigns provide two- to four-week average concentrations of NO\textsubscript{2} across 40 sampled locations as indicated in Fig. \ref{fig:pilot}a. Passive dosimeters results serve to validate the spatial patterns simulated by the urban air quality model described in subsection \ref{sec:cfd}. Table \ref{tab:sensors} summarizes the pollutants and meteorological variables measured by these different techniques, along with their temporal resolution.

Fig. \ref{fig:pilot}b compares the daily averaged concentrations from the co-located LCS and AQMS during June–August 2025. The results indicate that the low-cost sensors successfully captured the temporal variability of the reference measurements, showing slightly better performance for NO$_2$. Specifically, the LCS exhibited a mean bias of –7.24 $\mu g/m^3$ for NO$_2$ and –12.35 $\mu g/m^3$ for PM${10}$, with moderate correlation coefficients ($r$ = 0.58 for NO$_2$ and $r$ = 0.60 for PM${10}$). The corresponding root mean square errors were 10.48 $\mu g/m^3$ for NO$_2$ and 14.46 $\mu g/m^3$ for PM${10}$. These results are consistent with the expected accuracy of low-cost sensors.

\subsection{Edge computing: AI-based vehicle emission model and sensing data cleaning}
\begin{figure*}[ht]
    \centering
    \includegraphics[width=1\textwidth,trim= 7cm 2cm 7.0cm 0cm,clip]{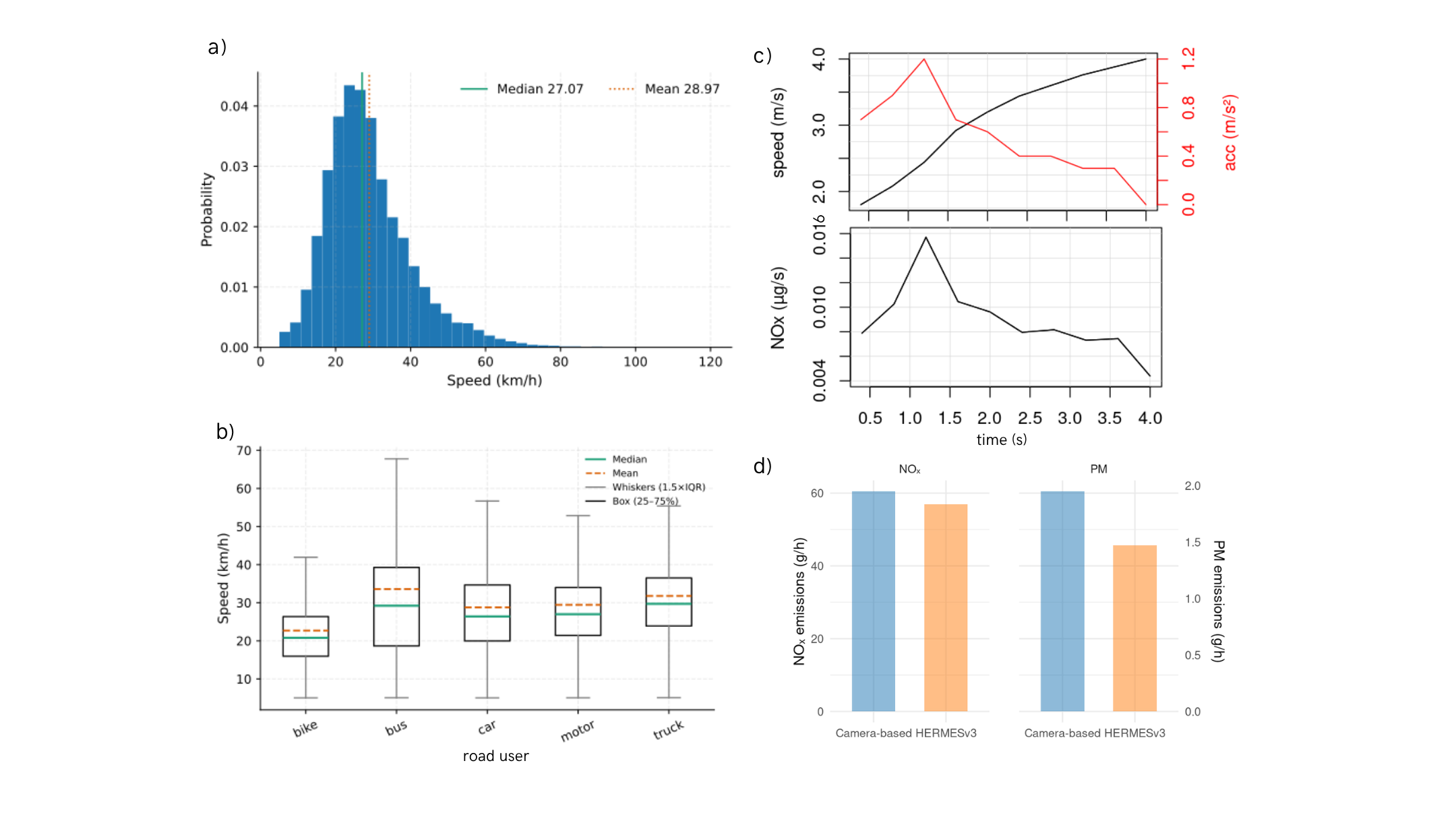}
    \caption{Vehicle detection and emission analysis derived from video-based tracking. (a) Probability density function of detected speeds for all road users. (b) Speed distribution by road user type. (c) Instantaneous speed, acceleration, and emission rates for an illustrative vehicle trajectory. (d) Comparison of one-hour aggregated emissions derived from camera-based estimates and from the HERMESv3 emission model based on average speed and expected vehicle counts.}
    \label{fig:camera}
\end{figure*}

In addition to the calibration analysis of LCS data, the key process executed at the edge is the derivation of real-time vehicular emissions. This is accomplished through two main steps. First, a dedicated software framework is designed to enable the seamless deployment and execution of real-time analytics at the edge, leveraging GPU resources and distributed computing paradigms to meet stringent latency requirements (\textit{Challenge 1}). Second, data analytics workflows are developed to extract relevant traffic information and estimate vehicular emissions, reducing the amount of data transmitted to the HPC environment and thereby facilitating parallel and scalable execution. (\textit{Challenge 2}).

For the software framework, \textbf{Kubernetes}\footnote{\url{https://kubernetes.io} accessed October 29, 2025} is adopted as the cloud-native execution environment, complemented by the \textbf{Prometheus}\footnote{\url{https://prometheus.io} accessed October 29, 2025} monitoring stack for collecting real-time performance metrics from the platform. The analytics are implemented using the \textbf{COMPSs} programming model and runtime framework \cite{PyCOMPSs2017}, which transforms sequential processes into distributed workflows. This enables the application’s inherent parallelism to be exploited by distributing tasks across the two available edge computing devices. The results of the analytics are stored locally through \textbf{MinIO}\footnote{\url{https://min.io} accessed October 29, 2025}, while \textbf{Apache Kafka}\footnote{\url{https://kafka.apache.org}  accessed October 29, 2025} is used to stream the processed data to the remote HPC infrastructure.

Regarding the AI-based video analytics, the following processes have been implemented: 

\begin{itemize}

    \item \textit{Step 1:} traffic surveillance cameras are physically connected to a central network router, which aggregates the incoming video streams and serves as the primary access point for nearby edge computing devices. These connections are established over Ethernet using the  Real Time Streaming Protocol (RSTP), a standard protocol designed for the efficient transmission and real-time control of multimedia streams. This networking setup enables low-latency and continuous video delivery between the cameras and the processing units, ensuring reliable performance for edge-based analysis.

    \item \textit{Step 2:} each incoming video frame is processed in real time by a GPU-enabled edge device running \textit{Camera-Edge}\footnote{\url{https://github.com/ProyectoAscender/camera-edge} accessed October 29, 2025}, a open source custom C++ application built upon the YOLOv6-M object detection framework\footnote{\url{https://github.com/meituan/YOLOv6} accessed October 31, 2025}. This module detects all visible road users in the frame and generates a corresponding set of \textit{bounding boxes}. Each box consists of: (i) a bounding rectangle defined by the pixel coordinates of its top-left corner, width and height, (ii) a predicted vehicle class (motorbike, passenger car, truck, or bus), and (iii) a confidence score quantifying the reliability of the detection. These metadata ensure that downstream components receive not only spatial location, but also semantic context and detection quality for each vehicle instance.     

    \item \textit{Step 3:} the set of detected vehicle boxes and their associated timestamps are transmitted via the User Datagram Protocol (UDP) to the \textit{Smart-City}\footnote{\url{https://github.com/ProyectoAscender/smart-city-compss} accessed October 31, 2025} process, an open-source tracking pipeline running at the edge to implement multi-object tracking and event detection. This process combines the ByteTrack algorithm \cite{bytetrack} with heuristic logic and a Kalman Filter to associate detections across frames and reconstruct complete vehicular trajectories. Once the trajectories are formed, each vehicle’s path is projected into Universal Transverse Mercator (UTM) coordinates, providing real-world spatial localization in meters. From these projected trajectories, instantaneous speeds are computed as finite differences in position over time. To reduce noise due to occasional detection inaccuracies, a rolling median filter is applied to smooth the resulting speed profiles. As a result, user road information can be derived such as the speed probability density function (Fig. \ref{fig:camera}a) and speed distribution as a function of road users (\ref{fig:camera}b). Furthermore, relevant traffic events of interest can also be detected, combining the position of the detected vehicles to predefined semantically annotated zones. 

    \item \textit{Step 4:} at each timestamp, the instantaneous Vehicle Specific Power (VSP) is calculated to characterize the vehicle’s power demand based on its speed, acceleration, and type. The VSP value directly link to emission factors in look-up tables that differentiate by vehicle category, fuel type, and Euro standard. These tables are pre-calculated from established micro-scale emission models; particularly \textit{PHEMlight}~\cite{hausberger2014extended} and SUMO~\cite{dlr124092}. For each detected vehicle, instantaneous emission rates of key pollutants (namely, NO$_x$ and PM) are estimated in real time. The emissions include both exhaust and non-exhaust components (cold-start effects, re-suspension, tire wear, and brake wear). We provide an illustrative example in Fig. \ref{fig:camera}c, in which speed, acceleration and the resulting NO$_x$ emission rates are plotted. These rates are then aggregated spatially by road segment and temporally over fixed intervals before being transmitted to the HPC infrastructure via cellular (4G/5G) connection.    
\end{itemize}

The aggregated real-time emissions have been compared against the High Elective Resolution Modelling Emission System version 3 (HERMESv3)\cite{guevara2020hermesv3}, currently used operationally by the Barcelona Supercomputing Center for air quality forecasting. HERMESv3 is a bottom-up model based on traffic flow data from the Barcelona City Council’s automatic counting network and average speed profiles from TomTom’s historical datasets. An illustrative example is shown in Fig. \ref{fig:camera}d, comparing hourly aggregated emissions during the morning traffic rush. For both NO$x$ and PM${10}$, camera-derived emissions were slightly higher than those estimated by HERMESv3. This difference arises because HERMESv3 does not capture individual vehicle dynamics, which are particularly important under stop-and-go conditions. In such situations, engines frequently operate at high load, producing short-term emission peaks and idling emissions during stops. As a result, camera-based estimates can exceed average-speed emission models, which tend to underestimate emissions in these traffic conditions.

Overall, the described analytics workflow enables real-time processing of video streams at the edge to derive aggregated vehicular emissions, which are subsequently transferred to HPC and cloud infrastructures for advanced air-quality modelling. This approach drastically reduces the volume of data transmitted (\textit{Challenge 2}), thereby minimizing latency and communication costs, while preserving privacy since only metadata, rather than raw video, are shared (\textit{Challenge 3}). Furthermore, the system is inherently scalable, leveraging cloud-native deployment automation and distributed-computing mechanisms to adapt seamlessly to different urban environments and computational resources.


\subsection{HPC computing: urban air quality modeling and data fusion}
\label{sec:cfd}
\begin{figure*}[ht]
    \centering
    \includegraphics[width=1\textwidth,trim= 2.1cm 0 0 0 cm,clip]{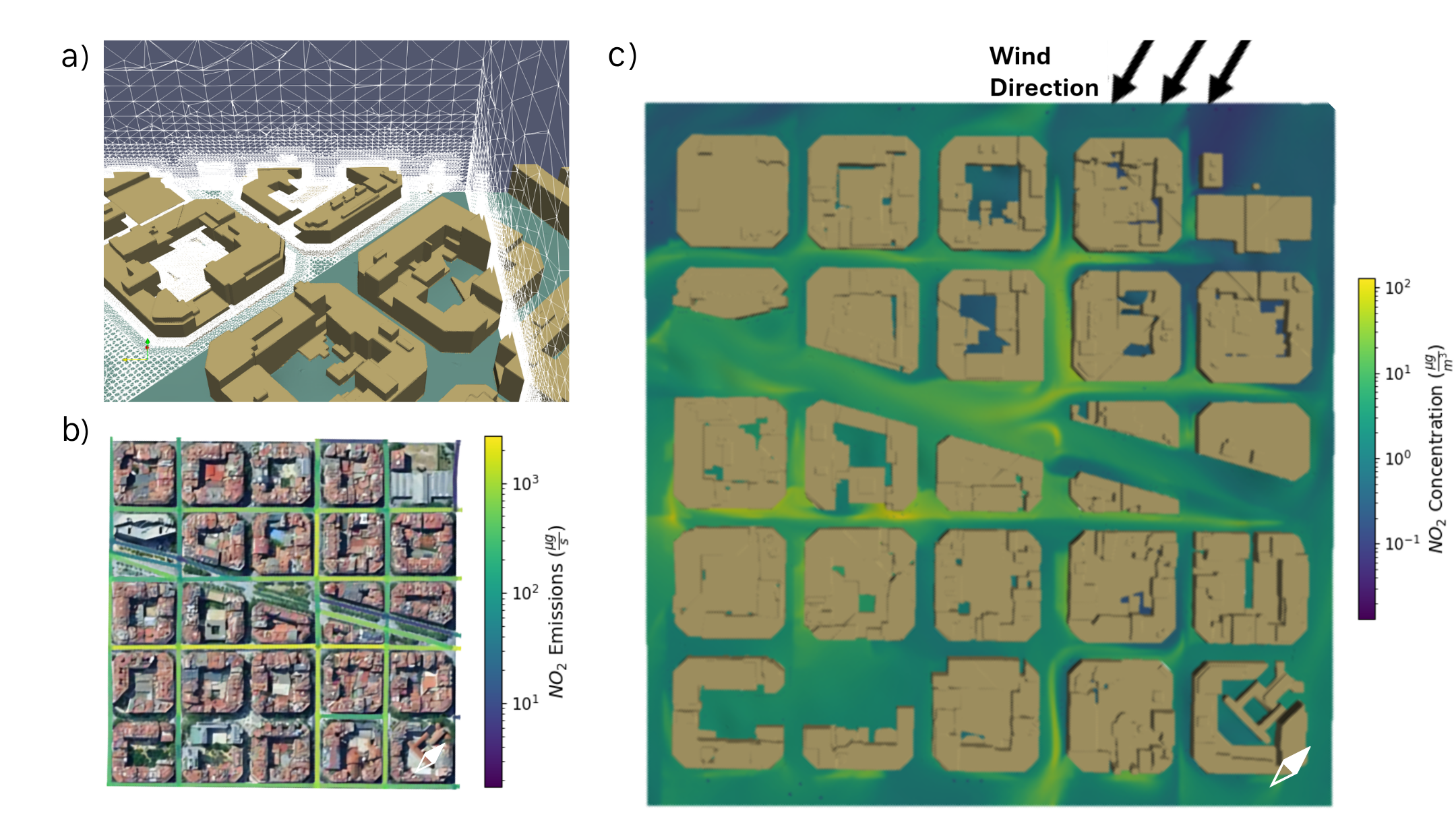}
    \caption{Example of a computational fluid dynamics simulation at the Barcelona pilot case. The panels show (a) details of the refined computational mesh, (b) a snapshot of NO\textsubscript{2} emissions for each road-link in the domain, and (c) the corresponding resulting NO\textsubscript{2} concentration field near ground level.}
    \label{fig:cfd}
\end{figure*}
The emission rates derived from the video-based tracking system can be used as input to advance in time the urban pollutant dispersion model. In the current pilot case, we use a high-resolution model based on computational fluid dynamics (CFD). The CFD simulations numerically integrate the governing transport equations for momentum and scalar transport using the Reynolds-Averaged Navier–Stokes (RANS) approach, implemented in the \textit{OpenFOAM} framework\footnote{\url{ www.openfoam.org}  accessed October 29, 2025}. The boundary conditions are extracted from the operational air quality system CALIOPE \cite{criado2023data}, which provides hourly data of the meteorological and concentration values at the mesoscale. The CFD computational domain was discretized into approximately 11 million cells using the \textit{SnappyHexMesh} tool in OpenFOAM. Local mesh refinement was applied near buildings and at ground level to better resolve velocity gradients and pollutant concentration variations within the urban canopy. Fig.~\ref{fig:cfd}a provides a close-up view of the mesh structure around buildings. As an example, Fig.~\ref{fig:cfd}b presents NO$_2$ emission estimates for a specific time step, derived from the HERMESv3 bottom-up emission model~\cite{guevara2020hermesv3}. The corresponding simulated concentration field, evaluated at 3~m above ground level, is shown in Fig.~\ref{fig:cfd}c.

Given the high computational cost of full CFD simulations, a surrogate modeling strategy was developed to enable real-time operation, addressing \textit{Challenge 4}. An extensive CFD database was constructed, covering a wide range of emission and meteorological scenarios. Then, the database was subsequently used for the training and validation of a surrogate model based on a clustering and a weighted-averaged strategy following previous work \cite{parra2010methodology}. The resulting surrogate reproduces CFD outputs at a fraction of the computational cost, enabling real-time execution and making it suitable for integration within the current operational system.

\begin{figure*}[ht]
    \centering
    \includegraphics[width=1\textwidth,trim= 2cm 3cm 2cm 6cm,clip]{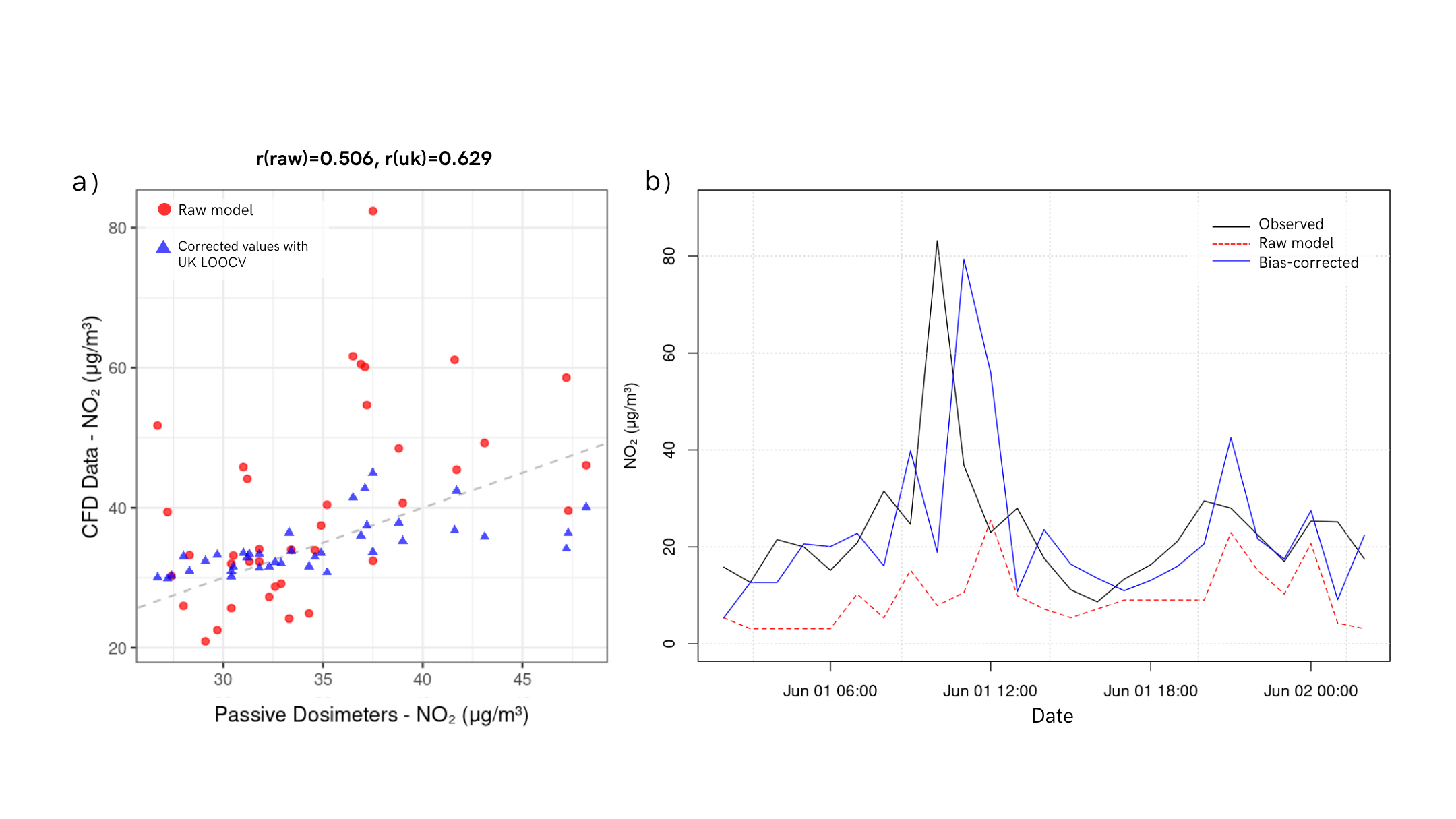}
    \caption{Model evaluation and data-fusion results. (a) NO$_2$ concentrations from dosimeters compared with raw CFD and Universal Kriging–corrected fields. (b) Temporal NO$_2$ evolution at the AQMS location for June 1, 2025, showing raw and Kalman Filter–corrected model predictions against observations.}
    \label{fig:validation}
\end{figure*}

An evaluation of long-term spatial patterns is presented in Fig. \ref{fig:validation}a, comparing NO$_2$ concentrations from a passive dosimeter campaign conducted between 18 September and 17 October 2024 with results from the CFD raw model. The sampling locations used for validation are shown in Fig. \ref{fig:pilot}a. Although the CFD simulation partially reproduces the observed NO$_2$ gradients, it remains affected by persistent uncertainties. Fig. \ref{fig:validation}b further illustrates the temporal evolution of NO$_2$ concentrations at the AQMS location compared with observations for June 1, 2025. Again, the raw CFD model presents notable uncertainties, partly due to simplifications introduced to reduce computational complexity, such as assuming isothermal flow, neglecting vegetation canopy effects, and treating pollutants as chemically inert, and partly due to uncertainties in boundary conditions and emission estimates. Nevertheless, the model successfully captures fine-scale urban flow structures and the influence of urban morphology on pollutant dispersion.

To illustrate the potential of data fusion techniques, we briefly present two approaches that combine observational data with model outputs. First, Universal Kriging (UK) is applied to bias-correct the model’s spatial patterns using data from the passive dosimeter campaign. UK is a well-established geostatistical method that combines a regression model with the spatial interpolation of its residuals. Fig. \ref{fig:validation}a compares leave-one-out cross-validation results from UK with those from the raw model. The bias-corrected fields show a substantial improvement (from a correlation coefficient of $r$=0.51 to $r$=0.63), demonstrating the capability of data fusion methods to enhance the characterization of pollution spatial patterns when dense networks of point measurements are available.

Second, a simple Kalman Filter is applied to correct the raw model-predicted time series shown in Fig. \ref{fig:validation}b. In this case, the Kalman Filter estimates the next model bias (model minus observation) based on the previously observed bias, enabling short-term corrections of model forecasts using real-time measurements. As illustrated in Fig. \ref{fig:validation}b, the bias-corrected results for June 1, 2025, show a notable improvement over the raw model, highlighting how even simple data assimilation techniques can significantly enhance air quality forecasts when real-time data are available.

\section{Open challenges and future research}
Significant research is still needed to fill the gap among IoT sensing networks, air quality models uncertainties, and requirements from urban planners and decision-makers. The following points outline the main open challenges where substantial progress is expected in the coming years.

\begin{itemize}

    \item \emph{Consistency and scalability in multi-camera systems: }Handling vehicle detection and tracking across overlapping camera views remains a challenge, as it can lead to emission overestimation. To mitigate this, dedicated deduplication algorithms should be put in place to improve the matching of vehicles through spatiotemporal correlation and feature similarity. Another key issue is determining fleet composition from camera data, since fuel type and Euro emission class are critical for accurate estimation but are rarely captured by tracking algorithms. In the current use case, these attributes are inferred from the average fleet composition of Barcelona urban area. A more precise approach would rely on automatic license plate recognition, but access to license plate databases is limited by privacy regulations and requires strict anonymization protocols.

    \item \emph{Human exposure in pollution hotspots:} Understanding how people move through polluted urban spaces is key to assessing the real health impacts of air quality. Combining human mobility patterns with concentration maps enables the identification of exposure, that is, when and where high pollutant concentrations coincide with high population density. IoT-based technologies, such as cameras coupled with computer vision algorithms, can provide valuable insights into the spatiotemporal distribution of pedestrian exposure across urban areas. This type of high-resolution, real-time information can support targeted and time-sensitive mitigation strategies. In our work in Barcelona, we demonstrated this approach by mapping pedestrian exposure in busy city corridors \cite{armengol2024city}.

    \item \emph{Quantification of model uncertainties:} Accurately quantifying and propagating uncertainties in both LCS and urban AQ models remains a critical challenge. On the modeling side, uncertainties arise from factors such as emissions, wind boundary conditions, background concentrations, urban canopy effects and geometric simplifications, and the inherent limitations of the CFD approach. More advanced techniques, such as large-eddy simulations, may provide higher-resolution insights to reduce some of these uncertainties. On the sensor side, LCS calibration can be improved using arrays of sensors to characterize cross-sensitivities, enabling compensation for interfering pollutants and environmental conditions, as detailed in \cite{Multisensor2020}. Systematically quantifying these uncertainties is particularly useful when integrating observations with model data. Over- or underestimating uncertainty can lead to misleading corrections in the model driven by noisy or biased observations. 

    \item \emph{Extending the study beyond road traffic:} The current pilot focuses on vehicular emissions as the dominant pollution source. However, the AI-based tracking and emission-estimation framework developed here can be adapted to other environments where different activities drive air pollution, such as maritime ports (ship emissions), airports (aircraft operations), or industrial and mining areas. Extending the methodology to these scenarios would require retraining detection models and adjusting emission factors to account for source-specific characteristics. Another promising direction is to identify and quantify the impact of traffic-related behaviors on pollution peaks, such as congestion, stop-and-go driving, or idling engines, as well as safety-relevant events (e.g., near-collisions inferred from predicted trajectories). These analyses would provide valuable insights into the relationship between traffic dynamics and pollution exposure, supporting more effective mobility policies and urban planning.

\end{itemize}

\section{Conclusions}

This work presents a multi-layer architecture for urban AQ monitoring that integrates IoT sensing with advanced analytic processes running across the compute continuum, from real-time edge computing to large-scale HPC simulations, with the cloud potentially being used for data storage, visualization, and digital-twin applications. Traffic cameras, low-cost sensors, and reference stations provide observational data that are processed at the edge to calibrate sensors and extract traffic emissions. These real-time observations feed HPC-based AQ simulations for both data assimilation and model validation. The deployment of an urban pilot in Barcelona demonstrates the feasibility of the approach and highlights the potential of combining IoT sensing with high-resolution urban AQ models to support more accurate and actionable assessments of pollution.

A key element of this architecture is the processing of video streams at the edge. By analyzing data locally, the system both preserves citizen privacy and drastically reduces the volume of information that needs to be transmitted, enabling the approach to scale to larger urban deployments without compromising performance or data protection. Deployment automation, distributed computing, and GPU acceleration make this possible, ensuring real-time efficiency, scalability, and privacy by design.

This integrated approach helps overcome the limitations of traditional monitoring and modelling. Physics-based AQ models offer full spatial coverage but still carry significant uncertainties, while point measurements alone do not capture broader urban patterns. By combining real-time observations with model simulations, the system delivers a more accurate and comprehensive picture of air quality.

Overall, this work demonstrates the potential of an IoT–edge–cloud–HPC continuum to provide high-resolution, privacy-conscious, and scalable urban air-quality monitoring, laying the groundwork for future research and operational applications in smart cities. The system can quantify the impact of traffic on air pollution in dense urban clusters, an aspect that is not easily captured with current methodologies. Moreover, there is a strong need to better assess the effects of new mobility and urban-planning policies adopted by many cities to reduce emissions, such as the creation of Low Emission Zones or new pedestrian areas with strict traffic limitations (e.g., the superblocks urban design in Barcelona or Distrito Zero in Madrid).

\section*{Acknowledgments}

This research has been funded by the AIR-URBAN project (TED2021-130210A-I00/ AEI/10.13039/501100011033/ European Union NextGenerationEU/PRTR). It is also financed by the ASCENDER project of the UNICO I+D Cloud program that has Ministry for Digital Transformation and of Civil Service and the EU-Next Generation EU as financing entities, within the framework of the PRTR and the MRR.

The authors would like to thank the Direcció General de Qualitat Ambiental i Canvi Climàtic – Generalitat de Catalunya for providing high-temporal-resolution observational data through the XVPCA. We also  thank 4sfera for supplying the experimental dosimeter campaign data and supporting fieldwork with low-cost sensors, and ARUP for their assistance in setting up the CFD simulations. Finally, we thank the BIT Habitat Foundation and the Barcelona City Council for enabling the urban pilot deployment.

\bibliographystyle{IEEEtran}
\bibliography{refs}

\begin{thebibliography}{10}
\providecommand{\url}[1]{#1}
\csname url@samestyle\endcsname
\providecommand{\newblock}{\relax}
\providecommand{\bibinfo}[2]{#2}
\providecommand{\BIBentrySTDinterwordspacing}{\spaceskip=0pt\relax}
\providecommand{\BIBentryALTinterwordstretchfactor}{4}
\providecommand{\BIBentryALTinterwordspacing}{\spaceskip=\fontdimen2\font plus
\BIBentryALTinterwordstretchfactor\fontdimen3\font minus \fontdimen4\font\relax}
\providecommand{\BIBforeignlanguage}[2]{{%
\expandafter\ifx\csname l@#1\endcsname\relax
\typeout{** WARNING: IEEEtran.bst: No hyphenation pattern has been}%
\typeout{** loaded for the language `#1'. Using the pattern for}%
\typeout{** the default language instead.}%
\else
\language=\csname l@#1\endcsname
\fi
#2}}
\providecommand{\BIBdecl}{\relax}
\BIBdecl

\bibitem{who2021who}
W.~H. Organization, \emph{WHO global air quality guidelines: particulate matter (PM2.5 and PM10), ozone, nitrogen dioxide, sulfur dioxide and carbon monoxide}.\hskip 1em plus 0.5em minus 0.4em\relax World Health Organization, 2021.

\bibitem{rodriguez2022extent}
D.~Rodriguez-Rey, M.~Guevara, M.~P. Linares, J.~Casanovas, J.~M. Armengol, J.~Benavides, A.~Soret, O.~Jorba, C.~Tena, and C.~P. Garc{\'\i}a-Pando, ``To what extent the traffic restriction policies applied in barcelona city can improve its air quality?'' \emph{Science of the Total Environment}, vol. 807, p. 150743, 2022.

\bibitem{guevara2020hermesv3}
M.~Guevara, C.~Tena, M.~Porquet, O.~Jorba, and C.~P{\'e}rez Garc{\'\i}a-Pando, ``Hermesv3, a stand-alone multi-scale atmospheric emission modelling framework--part 2: The bottom--up module,'' \emph{Geoscientific Model Development}, vol.~13, no.~3, pp. 873--903, 2020.

\bibitem{criado2023data}
A.~Criado, J.~M. Armengol, H.~Petetin, D.~Rodriguez-Rey, J.~Benavides, M.~Guevara, C.~P{\'e}rez Garc{\'\i}a-Pando, A.~Soret, and O.~Jorba, ``Data fusion uncertainty-enabled methods to map street-scale hourly no 2 in barcelona: a case study with caliope-urban v1. 0,'' \emph{Geoscientific Model Development}, vol.~16, no.~8, pp. 2193--2213, 2023.

\bibitem{rodulfo2020smart}
R.~Rodulfo, ``Smart city case study: City of coral gables leverages the internet of things to improve quality of life,'' \emph{IEEE Internet of Things Magazine}, vol.~3, no.~2, pp. 74--81, 2020.

\bibitem{cetinkaya2021distributed}
O.~Cetinkaya, B.~Zaghari, F.~M. Bulot, W.~Damaj, S.~A. Jubb, S.~Stein, A.~S. Weddell, M.~Mayfield, and S.~Beeby, ``Distributed sensing with low-cost mobile sensors toward a sustainable iot,'' \emph{IEEE Internet of Things Magazine}, vol.~4, no.~3, pp. 96--102, 2021.

\bibitem{mahajan2023ubiquitous}
P.~Mahajan, C.~R. Krishna, G.~S. Aujla, R.~S. Bali, and N.~Garg, ``Ubiquitous micro mapping of air pollution using vehicular edge-based connected iot services,'' \emph{IEEE Internet of Things Magazine}, vol.~5, no.~4, pp. 156--161, 2023.

\bibitem{martin2024using}
F.~Mart{\'\i}n, S.~Janssen, V.~Rodrigues, J.~Sousa, J.~L. Santiago, E.~Rivas, J.~Stocker, R.~Jackson, F.~Russo, M.~Villani \emph{et~al.}, ``Using dispersion models at microscale to assess long-term air pollution in urban hot spots: A fairmode joint intercomparison exercise for a case study in antwerp,'' \emph{Science of the Total Environment}, vol. 925, p. 171761, 2024.

\bibitem{PyCOMPSs2017}
E.~Tejedor, Y.~Becerra, G.~Alomar, A.~Queralt, R.~M. Badia, J.~Torres, T.~Cortes, and J.~Labarta, ``Pycompss: Parallel computational workflows in python,'' \emph{The International Journal of High Performance Computing Applications}, vol.~31, no.~1, pp. 66--82, 2017.

\bibitem{bytetrack}
\BIBentryALTinterwordspacing
Y.~Zhang, P.~Sun, Y.~Jiang, D.~Yu, Z.~Yuan, P.~Luo, W.~Liu, and X.~Wang, ``Bytetrack: Multi-object tracking by associating every detection box,'' \emph{CoRR}, vol. abs/2110.06864, 2021. [Online]. Available: \url{https://arxiv.org/abs/2110.06864}
\BIBentrySTDinterwordspacing

\bibitem{hausberger2014extended}
S.~Hausberger and D.~Krajzewicz, ``Extended simulation tool phem coupled to sumo with user guide,'' \emph{COLOMBO Project Report Deliverable}, vol.~4, 2014.

\bibitem{dlr124092}
\BIBentryALTinterwordspacing
P.~Alvarez~Lopez, M.~Behrisch, L.~Bieker-Walz, J.~Erdmann, Y.-P. Fl{\"o}tter{\"o}d, R.~Hilbrich, L.~L{\"u}cken, J.~Rummel, P.~Wagner, and E.~Wie{\ss}ner, ``Microscopic traffic simulation using sumo,'' in \emph{2005 IEEE Intelligent Transportation Systems Conference (ITSC)}.\hskip 1em plus 0.5em minus 0.4em\relax IEEE, November 2018. [Online]. Available: \url{https://elib.dlr.de/124092/}
\BIBentrySTDinterwordspacing

\bibitem{parra2010methodology}
M.~Parra, J.~Santiago, F.~Mart{\'\i}n, A.~Martilli, and J.~Santamar{\'\i}a, ``A methodology to urban air quality assessment during large time periods of winter using computational fluid dynamic models,'' \emph{Atmospheric Environment}, vol.~44, no.~17, pp. 2089--2097, 2010.

\bibitem{armengol2024city}
J.~M. Armengol, C.~Carnerero, C.~Rames, {\'A}.~Criado, J.~Borge-Holthoefer, A.~Soret, and A.~Sol{\'e}-Ribalta, ``City-scale assessment of pedestrian exposure to air pollution: A case study in barcelona,'' \emph{Urban Climate}, vol.~58, p. 102183, 2024.

\bibitem{Multisensor2020}
P.~Ferrer-Cid, J.~M. Barcelo-Ordinas, J.~Garcia-Vidal, A.~Ripoll, and M.~Viana, ``Multisensor data fusion calibration in iot air pollution platforms,'' \emph{IEEE Internet of Things Journal}, vol.~7, no.~4, pp. 3124--3132, 2020.

\end{thebibliography}
\newpage
\section{Biographies}
\begin{IEEEbiographynophoto}{Dr. Jan Mateu Armengol} is an Assistant Professor in the Fluid Mechanics Department at the Universitat Politècnica de Catalunya (UPC), a position he has held since 2023. He also leads the Air Quality Services team within the Earth Sciences Department at the Barcelona Supercomputing Center (BSC). Dr. Mateu Armengol holds a joint Ph.D. in Mechanical Engineering from the Universidade Estadual de Campinas (Brazil) and the Université Paris-Saclay (France) awarded in 2019. Jan’s research focuses on urban air quality modelling with emphasis in uncertainty quantification and data-fusion methods.
\end{IEEEbiographynophoto}
\begin{IEEEbiographynophoto}{Vicente Masip} is an industrial engineer with expertise in AI‐ML techniques and big data. He has worked in industry (2015‐2021) on software development, machine learning, and data analytics solutions with many frameworks (Python, R, Spark, Tensor Flow, etc.). Since 2021, he is a researcher in the PPC group at BSC. He participated in european projects such as CLASS, ELASTIC or EXTRACT, and is currently involved in ASCENDER and AIR URBAN.
\end{IEEEbiographynophoto}
\begin{IEEEbiographynophoto}{Ada Barrantes}  is a research engineer in the Air Quality Services team from the Earth Sciences Department at the Barcelona Supercomputing Center. She holds a Bachelor's degree in Physics and a Master's degree in Meteorology, both from the University of Barcelona (UB). Her work encompasses a range of tasks within the team, including programming, performing diagnostic analyses, data fusion, and organizing outreach activities.
\end{IEEEbiographynophoto}
\begin{IEEEbiographynophoto}{Gabriel Moreira Beltrami} has a degree in Mechanical Engineering from State University of Campinas (UNICAMP), General Engineering by Central School of Lyon (ECL) and a Master in Fluid Mechanics from ECL. Currently working in modeling pollutant concentration at very fine resolution in Barcelona, motivated by complex CFD problems and AI-based emulators methodologies.
\end{IEEEbiographynophoto}
\begin{IEEEbiographynophoto}{Sergi Albiach} holds a degree in Computer Science and a Master's degree in Artificial Intelligence. Since 2021, he has been working in the Predictable Parallel Computing (PPC) group, contributing to multiple projects, including DEEPHEALTH and AIR-URBAN among others. His research focuses on the integration of AI methodologies into complex computational systems through advanced machine learning methods and high-performance computing techniques.
\end{IEEEbiographynophoto}
\begin{IEEEbiographynophoto}{Dr. Daniel Rodriguez } holds a MSc in Air Quality by the University of Birmingham and a B.S. in Chemical Engineering from the Polytechnic University of Catalonia, where he also obtained his Ph.D in Environmental Engineering in 2022. His expertise is focused on urban air quality and the application of traffic simulators to air quality systems. He is currently working as a researcher at the Energy Climate and Urban transition unit in TECNALIA focused in the development of air quality tools based on AI. 
\end{IEEEbiographynophoto}
\begin{IEEEbiographynophoto}{Dr. Albert Soret}, PhD in Environmental Engineering from the Polytechnic University of Catalonia, is head of the Earth System Services group at BSC and an adjunct lecturer at the Polytechnic University of Catalonia. With 15 years of experience in air quality and climate research, his expertise spans atmospheric emissions, meteorological and air quality modeling, and climate services. He leads multiple national and international projects (e.g., ASPECT, S2S4E, Clim4Energy, VISCA, AI4DROUGHT) and coordinates the CALIOPE Air Quality Forecast System for Spain and Catalonia. His work emphasizes technology transfer for sustainable development across urban development, energy, transport, health, agriculture, and water sectors.
\end{IEEEbiographynophoto}
\begin{IEEEbiographynophoto}{Dr. Marc Guevara}, B.S. in Industrial Engineering and PhD in Environmental Engineering from the Polytechnic University of Catalonia. He is a team leader and established researcher with 15 years’ experience in the areas of emission and air quality modelling. He coordinates the scientific development of the BSC in-house HERMES emission model. He is co-chair of the Emissions Working Group of the FAIRMODE community and member of the GEIA Scientific Steering Committee.  He has participated in multiple European H2020 and Horizon Europe projects (AQ-WATCH, CoCO2, CORSO, CAMEO, MARCHES, RI-URBANS) and Copernicus contracts (CAMS$\_$81, CAMS$\_$COP066, CAMS$\_$COP079, CAMS2$\_$61, CAMS2$\_$40).
\end{IEEEbiographynophoto}
\begin{IEEEbiographynophoto}{Dr. Eduardo Quiñones}is a senior researcher and head of the PPC group at BSC, with consolidated expertise in distributed and parallel computation for critical real-time systems. He has coordinated four EU-funded projects (EXTRACT, AMPERE, ELASTIC, CLASS). His research interests focus on the applicability of distributed and parallel processing for AI in the health, automotive, space, and smart agriculture domains.
\end{IEEEbiographynophoto}
\begin{IEEEbiographynophoto} {Dr. Elli Kartsakli} is an established researcher at BSC with R3 accreditation and activity leader of the Edge to Cloud Computing Paradigms area within the Predictable Parallel Computing group. She is a recipient of a Ramon y Cajal 5-year fellowship and specializes in edge-to-cloud computing architectures and orchestration paradigms, leveraging 5G and beyond communications. She holds a PhD in Telecommunications and has experience in both academia and industry. She has extensive expertise in research and development projects, focusing on smart-city applications and healthcare (VERGE, ELIXIRION, AIR-URBAN, PROXIMITY, FORGE).
\end{IEEEbiographynophoto}
\end{document}